\documentclass[conference]{IEEEtran}
\IEEEoverridecommandlockouts
\usepackage{cite}
\usepackage{amsmath,amssymb,amsfonts}
\usepackage{algorithmic}
\usepackage{algorithm}
\usepackage{graphicx}
\usepackage{verbatim}
\usepackage{textcomp}
\usepackage{xcolor}
\usepackage{url}
\usepackage{bm}
\usepackage{subfigure}
\usepackage{flushend}
\renewcommand{\algorithmicrequire}{\textbf{Initialize:}}

\def\BibTeX{{\rm B\kern-.05em{\sc i\kern-.025em b}\kern-.08em
    T\kern-.1667em\lower.7ex\hbox{E}\kern-.125emX}}
\begin{document}

\title{Energy Efficient Transmission Parameters Selection Method Using Reinforcement Learning \\in Distributed LoRa Networks
}
\author{Ryotai Airiyoshi\IEEEauthorrefmark{1},
Mikio Hasegawa\IEEEauthorrefmark{2}, 
Tomoaki Ohtsuki\IEEEauthorrefmark{3},
and Aohan Li\IEEEauthorrefmark{1}
\\
\IEEEauthorrefmark{1}Graduate School of Informatics and Engineering, The University of Electro-Communications, Tokyo, Japan\\
\IEEEauthorrefmark{2}Department of Electrical Engineering, Tokyo University of Science, Tokyo, Japan\\
\IEEEauthorrefmark{3}Department of Information and Computer Science, Keio University, Yokohama, Japan\\
a2431010@gl.cc.uec.ac.jp, hasegawa@ee.kagu.tus.ac.jp, 
ohtsuki@keio.jp,
aohanli@ieee.org
}

\maketitle

\begin{abstract}
With the increase in demand for Internet of Things (IoT) applications, the number of IoT devices has drastically grown, making spectrum resources seriously insufficient. Transmission collisions and retransmissions increase power consumption. Therefore, even in long-range (LoRa) networks, selecting appropriate transmission parameters, such as channel and transmission power, is essential to improve energy efficiency. However, due to the limited computational ability and memory, traditional transmission parameter selection methods for LoRa networks are challenging to implement on LoRa devices. To solve this problem, a distributed reinforcement learning-based channel and transmission power selection method is proposed, which can be implemented on the LoRa devices to improve energy efficiency in this paper. Specifically, the channel and transmission power selection problem in LoRa networks is first mapped to the multi-armed-bandit (MAB) problem. Then, an MAB-based method is introduced to solve the formulated transmission parameter selection problem based on the acknowledgment (ACK) packet and the power consumption for data transmission of the LoRa device. The performance of the proposed method is evaluated by the constructed actual LoRa network. Experimental results show that the proposed method performs better than fixed assignment, adaptive data rate low-complexity (ADR-Lite), and $\epsilon$-greedy-based methods in terms of both transmission success rate and energy efficiency.
\end{abstract}


\section{Introduction}
\label{sect:introduction}
The Internet of Things (IoT) has gained attention as a mechanism for connecting and controlling objects via the Internet. Consequently, the number of IoT devices is rapidly increasing and is expected to reach 44 billion by 2025 \cite{A1}. As a result, communication resource shortages are expected to become more severe. This could lead to network delays and increased power consumption due to channel congestion, interference, and retransmissions. Additionally, it is challenging to recharge a large amount of IoT devices frequently. To address these issues, techniques that can efficiently use wireless resources while consuming low power is necessary.

Wireless communication technologies such as Bluetooth, Zigbee, Wi-Fi, and 4G, which have been used for conventional IoT, have limitations such as high power consumption, short transmission distance, and high transmission overhead. On the other hand, Low Power Wide Area Networking (LPWAN) is gaining attention due to its low power consumption, long-distance communication, low transmission overhead, and low bit rate support \cite{A2}. Long Range (LoRa) WAN is one of the representative LPWAN techniques. Compared to other LPWAN techniques, such as SIGFOX, it is open source, which allows the construction of autonomous networks at a low cost without relying on communication infrastructure. In addition, LoRa devices have low power consumption due to infrequent communication, allowing several years of battery life. Additionally, LoRa offers high payload capacity (up to 250 Bytes), and communication ranges from 2 km to 15 km, providing better downlink functionality than Sigfox or Ingenu \cite{A2}.
Furthermore, LoRa operates in the unlicensed ISM band, so no radio license is required \cite{A3}. 

Due to the superiority of the LoRa technique, this paper focuses on the LoRa networks.
Pure Aloha is the default media access protocol for LoRa networks, which is simple and has low overhead, resulting in low power consumption \cite{B1}. 
However, as the number of LoRa devices increases, Pure Aloha may face a high probability of communication collision, which may reduce the spectrum and energy efficiency. 
Therefore, selecting suitable communication parameters dynamically to improve the spectrum and energy efficiency is crucial.


There are two types of communication parameter selection methods in LoRa networks: centralized and distributed. Centralized methods include the Adaptive Data Rate Low-complexity scheme (ADR-Lite) \cite{A13} and Low-Power Multi-Armed Bandit (LP-MAB) \cite{A12}. ADR-Lite is a low-complexity algorithm that determines transmission parameters without considering previous packet history. However, it faces the challenge of increasing energy consumption in high-density networks, as it does not avoid choosing higher transmission power to increase transmission success rates \cite{A13}. LP-MAB is an algorithm that improves energy consumption by centrally setting end-device transmission parameters at the network server side, and simulations have shown high packet delivery ratio (PDR) and low energy consumption \cite{A12}. However, this centralized method may increase latency and end-device power consumption because the gateway (GW) selects parameters and informs the LoRa device which channel to access. Meanwhile, the LoRa devices need to listen to the information transmitted from the GW.

On the other hand, distributed methods \cite{A5, C1, C2,C6} include the cooperative multi-agent deep reinforcement learning prioritized experience replay (DRL-PER) \cite{A5}, and the MAB algorithm-based \cite{C1, C2,C6} methods. Cooperative multi-agent DRL-PER is a communication resource management approach that uses deep reinforcement learning to achieve high transmission success rates and energy efficiency. However, it requires a large amount of computation and memory, making it unsuitable for resource-constrained IoT devices \cite{A5}. The MAB algorithm is a lightweight and can be operated with minor computational power and memory and exhibits reasonable transmission success rates while maintaining fairness among devices \cite{ C1, C2, C6, C3, C4, C5}. However, energy efficiency is not considered in the related work \cite{ C1, C2, C6}. 



As described above, related work on centralized methods involves the network server acting as an agent to set transmission parameters, which may increase the consumption of communication resources, and they have yet to be verified on real devices \cite{A12, A13}. On the other hand, decentralized methods either do not consider energy efficiency or require significant computation and memory, making them difficult to apply to resource-constrained IoT devices \cite{A5, C1, C2,C6}. 
To address these challenges, we propose an implementable transmission parameter selection method that considers energy efficiency for resource-constrained IoT. 
The main contributions of this paper are summarized as follows.


\begin{itemize}
      \item We propose an autonomous decentralized lightweight reinforcement learning-based transmission parameter selection method to improve energy efficiency for LoRa networks. In the proposed method, the channel and transmission power are selected using ACK information and the magnitude of the selected transmission power. This method can be implemented on IoT devices with small memory capacity and low computational ability.
      \item By considering the magnitude of power consumption in the reward part of the proposed method, lower transmission power can be selected while maintaining communication reliability, thus improving energy efficiency as well as the transmission success rate.
      \item The proposed method is implemented on actual LoRa devices, and its performance is evaluated using the constructed actual high-density LoRa network. Experimental results show the superiority of the proposed method in terms of energy efficiency and transmission success rate in the real environment.
\end{itemize}

This paper is structured as follows. Section II introduces the system model and problem formulation. Section III describes the proposed method. Section IV presents the performance evaluation. Section V concludes this paper.

\section{SYSTEM MODEL}
\label{sect:systemmodel}
This section describes the system model, followed by the energy consumption model and problem formualtion. 

\subsection{System Model}

As shown in Fig. 1, this paper considers a LoRa network consisting of one gateway (GW) and $N$ LoRa end devices (EDs). The set of EDs is denoted by $\mathcal{N}$. 
We define the set of available channels as $\mathcal{C} = \{ c_1, c_2, \dots,c_m, \dots, c_M \}$ and that of transmission powers as $\mathcal{K} = \{ k_1, k_2, \dots,k_p, \dots, k_P \}$, where $M$ and $P$ are the number of channels and that of transmission power level, respectively, $c_m$ and $k_p$ represent the $m$-th channel and the $p$-th transmission power (TP) level, respectively. Each LoRa ED selects one channel from the set $\mathcal{C}$ and one TP level from the set $\mathcal{K}$ before transmitting.

\begin{figure}
\centering
\includegraphics[width=8cm]{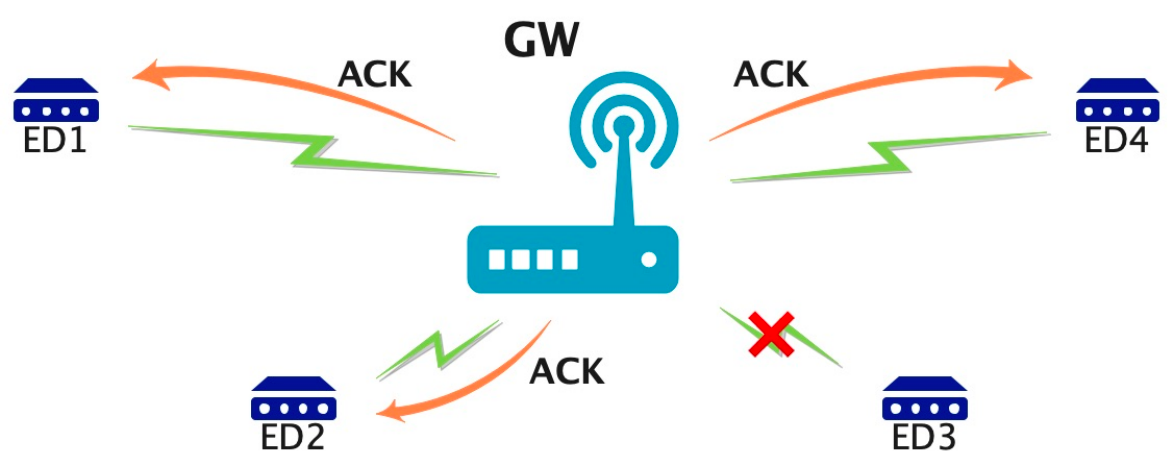}
\caption{System Model}
\label{fig:s1}
\end{figure}

Each LoRa ED sends data to the GW at regular intervals, performs carrier sense to check if the selected channel is available before transmitting, and transmits data to the GW if the channel is available. When the GW receives data from the ED successfully, it returns an ACK information. If the transmission fails, as with ED3 in Fig. \ref{fig:s1}, no ACK information is sent from the GW.
The transmission parameters selection algorithm is implemented on each LoRa ED to learn to select the transmission parameters, e.g., channel and TP level, using ACK information returned from the GW and the power consumption after each time of data transmission. 

\subsection{Energy Consumption Model}
The energy consumption model considered in this work is the energy consumption during data communication in active mode \cite{A12}. The formula is shown below.
\begin{itemize}
    \item \textbf{Energy Consumption in Active Mode ($E_{Active}$)}
    \begin{equation}
      E_{Active} = E_{WU} + E_{\text{proc}} +  E_{\text{ToA}} + E_{\text{R}},
    \end{equation}   
   where \(E_{\text{WU}}\) is the energy consumption for starting up the LoRa ED, \(E_{\text{proc}}\) is the energy consumption of the microcontroller's processing for selecting the transmission parameters, \(E_{\text{ToA}}\) is the energy consumption during data transmission, and \(E_{\text{R}}\) is the energy consumption in the receive mode of the LoRa ED. 
   $E_{\text{WU}}$, \(E_{\text{proc}}\), and \(E_{\text{R}}\) depend on the specifications of the modules used in the device.
    \item \textbf{Energy Consumption During Data Transmission (\(E_{\text{ToA}}\))}
    \begin{equation}
       E_{\text{ToA}} = (P_{\text{MCU}} + P_{\text{ToA}}) \cdot T_{\text{ToA}},
       \label{EToA}
    \end{equation}  
    where \(P_{\text{MCU}}\) is the energy consumption of the microcontroller during startup, \(P_{\text{ToA}}\) is the energy consumption rate during data transmission, and \(T_{\text{ToA}}\) is the total time taken for data transmission. \(P_{\text{ToA}}\) is determined by the selected transmission power.
    \item \textbf{Calculation of Time-On-Air (\(T_{\text{ToA}}\))}
    \begin{equation}
      T_{\text{ToA}} = T_{\text{Preamble}} + T_{\text{Payload}},
    \end{equation}
    where \(T_{\text{Preamble}}\) is the time for transmitting the preamble, and \(T_{\text{Payload}}\) is the time taken to transmit the data payload.
    \item \textbf{Calculation of Preamble Time (\(T_{\text{Preamble}}\))}
    \begin{equation}
      T_{\text{Preamble}} = (4.25 + N_{\text{P}}) \cdot T_{\text{Symbol}},
    \end{equation}
    where \(N_{\text{P}}\) is the number of preamble symbols, and \(T_{\text{Symbol}}\) is the transmission time for each symbol.
    \item \textbf{Calculation of Payload Time (\(T_{\text{Payload}}\))}
    \begin{equation}
      T_{\text{Payload}} = T_{\text{Symbol}} \cdot N_{\text{Payload}},
    \end{equation}
    where \(N_{\text{Payload}}\) is the number of payload symbols.
   \item \textbf{Calculation of Symbol Time (\(T_{\text{Symbol}}\))}
    \begin{equation}
      T_{\text{Symbol}} = \frac{2^{SF}}{BW},
    \end{equation}
    where \(SF\) is the spreading factor, and \(BW\) is the bandwidth.
\end{itemize}

\subsection{Problem Formulation}
This paper aims to maximize the energy efficiency of LoRa EDs by using reinforcement learning to select the optimal channel and the smallest TP level. The object function considered in this paper is as follows.

\begin{equation}
(P)=\underset{k_p \in \mathcal{K},{c_m \in \mathcal{C}}}{\max}\sum_{t=1}^{T}{EE}_{(k_p,c_m)}(t),
\end{equation}
where $T$ represents the maximum number of transmission attempts, and ${EE}_{(k_p,c_m)}(t)$ is defined as:
\begin{equation}
{EE}_{(k_p,c_m)}(t)=\frac{X_{(k_p,c_m)}(t)}{E_{Active}},
\end{equation}
where $X_{(k_p,c_m)}(t)$ is the  transmission success rate for the transmission parameters combination $(k_p,c_m)$ at time $t$ \cite{C3}, which is defined as: 
\begin{equation}
  X_{(k_p,c_m)}(t)=\frac{R_{(k_p,c_m)}(t)}{N_{(k_p,c_m)}(t)},
\end{equation}
where \(R_{(k_p,c_m)}(t)\) and \(N_{(k_p,c_m)}(t)\) are the number of successful transmissions and that of the selections of parameter set \((k_p,c_m)\) at time $t$, respectively.

\section{Proposed Method}
\label{sec:sec3}
This section introduces our proposed method. First, we describe the MAB problem and one of the representative MAB algorithms, i.e., UCB1-tuned. Then, the proposed method based on the UCB1-tunned of this study is presented.

\subsection{What is the MAB Problem?}
The MAB problem involves an agent (player) selecting and playing among multiple slot machines \cite{A9}\cite{A10}. The player aims to maximize the reward (coins) obtained by playing repeatedly. The player learns the probability of coins appearing for each slot machine through repeated play. The player should play and explore many slot machines to find the most coin-giving slot machine. However, excessive exploration can prevent maximizing coin acquisition. Thus, the MAB problem is a decision-making problem considering the trade-off between exploring slot machines to find the most coin-giving one and exploiting good slot machines to increase coins.\\

\subsection{MAB Algorithm}
The MAB algorithm is designed to solve the MAB problem. The transmission parameters selection method proposed in this paper is based on the UCB1-tuned, representative MAB algorithm. The UCB1-tunned algorithm is briefly explained below.

Auer and Bianchi proposed the UCB series of algorithms \cite{A11}. UCB1 is a simple reinforcement learning algorithm that balances exploitation and exploration by considering the average reward obtained and the total number of selections made until the present. The UCB1-tuned algorithm enhances the UCB1 algorithm, which considers the variance in the empirical rule of each slot machine. The specific operation involves playing each slot machine once initially and then selecting the machine to play according to the following equation at the \(t\)-th play.\\
\begin{equation}
  u^*_i =\underset{u_i \in \mathcal{U}} {\operatorname{argmax}}P_{u_i}(t),
  \label{UCBscore1}
\end{equation}
where \(P_{u_i}(t)\) is the UCB score of slot machine \(u_i\) at the \(t\)-th play, which is expressed as follow.
\begin{equation}
  P_{u_i}(t) = \frac{R_{u_i}(t)}{N_{u_i}(t)} + \sqrt{\frac{\ln t}{N_{u_i}(t)} \min\left(\frac{1}{4}, V_{u_i}(t)\right)},
  \label{UCBscore}
\end{equation}
where $R_{u_i}(t)$ and $N_{u_i}(t)$ are the sum of the rewards and the number of times taken of the slot machine $u_i$ prior to time $t$, respectively. $V_{u_i}(t)$ is the variance estimator that takes into account the number of plays $t$ and the total number of the selections of slot machine $u_i$, which can be expressed as:
\begin{equation}
  V_{u_i}(t) = \sigma^2_{u_i} + \sqrt{\frac{2 \ln t}{N_{u_i}(t)}},
\end{equation}
where $\sigma^2_{u_i}$ represents the variance of the received rewards of the slot machine $u_i$.

\subsection{Transmission Parameter Selection Method Using the UCB1-Tuned Algorithm}


A key feature of the proposed method is its combination of autonomy and lightweight design. Each ED operates autonomously, learning the communication environment and selects appropriate transmission parameters without relying on centralized control. Additionally, because the learning process is based solely on ACK information and TP, the amount of stored information is small. This lightweight design ensures the proposed method can be easily implemented on IoT devices with limited computational and memory resources. Furthermore, each ED learns and prioritizes less congested channels and optimal transmission power levels utilizing ACK information, the channel congestion and communication collisions can be effectively mitigated. 

Specifically, in our proposed method, each LoRa ED, based on the implemented UCB1-tuned algorithm, selects the channel and TP level. 
EDs wait for ACK information from the GW and determine rewards based on the presence or absence of this information. The reward for receiving ACK information is defined as 1/$E_{\text{ToA}}$. Otherwise, the reward is defined as 0. Thus, by changing the reward to be given according to the magnitude of power consumption, a smaller transmission power can be selected. Each ED can calculate its energy efficiency after transmission using the selected transmission parameters (channel and TP level). 
Each ED starts operation at random timings and repeats the selection of the transmission parameters and transmission using the selected parameters at regular intervals.
Based on the description above, to make it easy to understand, we summarize the relationship between the MAB problem and the transmission parameters selection problem in this work shown in Table \ref{tab:MAB} while the proposed method is summarized in Algorithm 1.

\begin{table}
  \centering
  \caption{Comparison between MAB Problem and Transmission Parameters Selection Problem}
  \label{tab:MAB}
  \begin{tabular}{{c|c}}
    \hline
    \textbf{MAB Problem} & \textbf{Channel, TP Selection Problem} \\
    \hline
    Player & LoRa ED \\
    \hline
    Slot Machine & Channel, TP \\
    \hline
    Reward: Coins & Reward: ACK information and $E_{ToA}$  \\
    \hline
    Objective: Maximize Coins & Objective: Maximize Energy Efficiency \\
    \hline
  \end{tabular}
\end{table}

\begin{algorithm}
\small
\caption{Proposed Method}
\begin{algorithmic}[1]
\renewcommand{\algorithmicrequire}{\textbf{Initialize:}}
\REQUIRE $t = 0$, ${R_{(k_p,c_m)}(t)} = 0$, ${N_{(k_p,c_m)}(t)} = 0$, $V_{(k_p,c_m)}(t) = 0$

\FORALL{channel and TP combinations}
    \STATE Try transmission
    \STATE $t = t + 1$
\ENDFOR

\WHILE{$t \leq T$}
    \STATE Select the channel from the channel set $\mathcal{C}$ and TP level form the TP set $\mathcal{K}$ with the maximum UCB score in energy efficiency, respectively.
    \STATE Packet transmission use the selected channel and TP
    \STATE Check whether an ACK frame has been received for the packet transmission
    \STATE Calculate power consumption $E_{\text{ToA}}$ using (\ref{EToA})
    \STATE Update UCB score using (\ref{UCBscore}) in energy efficiency
    \STATE $t = t + 1$
    \STATE Sleep mode
\ENDWHILE
\end{algorithmic}
\end{algorithm}

In the proposed method, all variables are initialized as 0 first. Then, each LoRa ED transmits once using each channel and TP level combination (lines 1-4 in Algorithm 1). After that, the channel and TP level with the highest value calculated based on (\ref{UCBscore}) are selected, and data are transmitted using the selected parameters (lines 6 and 7 in Algorithm 1). After transmission, the ACK information returned from the GW is checked, and the UCB score is updated based on the presence or absence of ACK information and the level of the selected TP (lines 8-10 in Algorithm 1). This operation is repeated a certain number of times, i.e., $T$. Here, $t$ is the number of transmissions.

\section{Performance Evaluation}
\label{sect:simulation result}
In this section, we first describe the experimental environment and parameter settings. Then, the compared algorithms are introduced. Finally, the selected ratio of each transmission power level, the performance in transmission success rate, and energy efficiency are evaluated and compared.

\begin{figure}
\centering
\includegraphics[width=8cm]{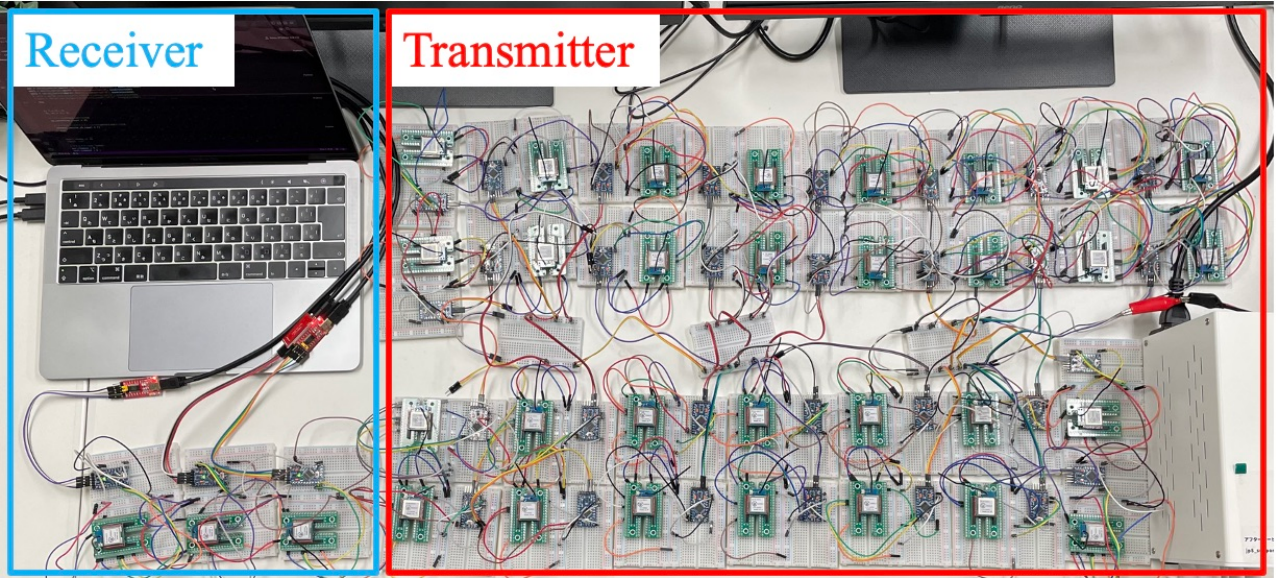}
\caption{Experimental Environment}
\label{fig:kankyo}
\end{figure}

\subsection{Experimental Environment and Parameter Settings}
In this work, LoRa EDs implementing the proposed method were set up as shown in Fig. \ref{fig:kankyo}. 
The transmitter and receiver sides consist of the LoRa EDs and the GW, respectively. The transmitters generate and transmit data. The receiver processes the received data to observe the transmission success rate and energy efficiency of each transmitter. Three receivers were set up to imitate GW, and each was assigned a different received channel. The number of LoRa EDs varies from 10 to 30 in the performance evaluations. 
Each device starts operation at random timings and transmits data every 10 seconds. The bandwidth (BW) and SF are set to 125 kHz and 7, respectively. The number of retransmissions is set to 0. Each LoRa ED selects one channel from five channels, i.e., \{920.6, 921.0, 921.4, 921.8, 922.2\} MHz, and one transmission power from the transmission power set \{-3, 1, 5, 9, 13\} dBm. The GW can only receive the transmitted data using the following three channels: \{921.0, 921.4, 921.8\} MHz. The transmission times for each LoRa ED is set to 200. Parameters related to the energy consumption mode used in the performance evaluation are $E_{WU}=(56.1 * T_{\text{WU}})$ mWh, $E_{proc}=(85.8 * T_{\text{proc}})$ mWh, $E_{R}=(66 * T_{\text{R}}$) mWh, $P_{MCU}=29.7$ mW, $N_{P}=8$bytes, $N_{Payload}=36\sim44$ bytes. Here, $T_{\text{WU}}$ represents the wake-up time of the LoRa device, $T_{\text{proc}}$ is the processing time for selecting transmission parameters by the microcontroller, and $T_{\text{R}}$ is the reception time of the device. 
The details of the experimental parameters are summarized in Table \ref{tab:para}.

\begin{table}
  \centering
  \caption{Parameter Settings in Experiment}
  \label{tab:para}
  \begin{tabular}{{c|c}}
    \hline
    Parameter & Value \\
    \hline
    \hline
    Number of Devices & 10, 15, 20, 25, 30 \\
    \hline
    BW & 125 kHz \\
    \hline
    SF & 7 \\
    \hline
    Selectable channel & 920.6, 921.0, 921.4, 921.8, 922.2 MHz \\
    \hline
    Receivable channel & 921.0, 921.4, 921.8 MHz \\
    \hline
    Selectable TP & -3, 1, 5, 9, 13 dBm\\
    \hline
    Transmission Interval & 10 s \\
    \hline
    Number of Retransmissions & 0 \\
    \hline
    Number of Transmissions & 200 times \\
    \hline
    $E_{WU}$ & $56.1 * T_{\text{WU}}$ mWh \\
    \hline
    $E_{\text{proc}}$ & $85.8 * T_{\text{proc}}$ mWh \\
    \hline
    $E_{\text{R}}$ & $66 * T_{\text{R}}$ mWh \\
    \hline
    $P_{\text{MCU}}$ &  29.7 mW \\
    \hline
    $N_{\text{Payload}}$ & 36 $\sim$ 44 bytes \\
    \hline
    $N_{\text{P}}$ & 8 bytes \\
    \hline
  \end{tabular}
\end{table}

\subsection{Comparison Methods}
To examine the effectiveness of the proposed method, we compare our proposed method with the $\epsilon$-greedy-based, ADR-Lite, and fixed transmission parameters allocation methods.
$\epsilon$-greedy is the simplest MAB algorithm, where LoRa EDs select the combination of the channel and transmission power with the largest reward with probability $1-\epsilon$ and randomly select a combination with probability $\epsilon$.
The fixed allocation method pre-assigns CHs evenly to transmitters and transmits at the minimum TP. 
The ADR-Lite algorithm was introduced as a centralized method, but it is implemented as a distributed method in this performance evaluation. In the ADR-Lite algorithm, the LoRa ED sorts the transmission power in an increased order while the channel is listed according to the channel situation.
Specifically, the list of the transmission parameters is set as: \{\{CH1, -3 dBm\}, \{CH9, -3 dBm\}, \{CH3, -3 dBm\}, \{CH5, -3 dBm\}, \{CH7, -3 dBm\}, \{CH1, 1 dBm\}, {CH9, 1 dBm\}, \{CH3, 1 dBm\}, \dots, \{{CH1, 13 dBm\}, \{CH9, 3 dBm\}, \{CH3, 13 dBm\}, \{CH5, 13 dBm\}, \{CH7, 13 dBm\}\}, where CH1, CH3, CH5, CH7, and CH9 are the channels with 920.6, 921.0, 921.4, 921.8, and 922.2 MHz, respectively. CH1 and CH9 are unavailable for the receiver, which can be regarded as the channels with the worst situation. The combination of the channel and TP located further back of the transmission parameter list, the TP is higher while the channel situation is better. In the ADR-Lite algorithm, LoRa ED initiates communication starting with the last combination of the transmission parameters in the list first. If the transmission is successful, the next set of the transmission parameters is halved to the middle value of the first set and the previously selected transmission parameter set in the list; if it fails, the next set of the transmission parameters is set to the transmission parameters in the middle of the last set, and the previously selected transmission parameter set in the list.

\subsection{Experimental Results}
In this subsection, we evaluate the selection ratio of each transmission value, transmission success rate, and the energy efficiency of the proposed method. For each result, it was the average value for five times of experiments.

\subsubsection{Selection ratio of each TP level}
Fig. \ref{fig:TPRatio} shows the proportion of transmission power selected when the transmission was successful for each algorithm in the experiment conducted with 30 transmitters. Fixed allocation is omitted in the results because it transmits at the minimum TP for all transmissions.
\begin{figure}
\centering
\includegraphics[width=50mm, angle=270]{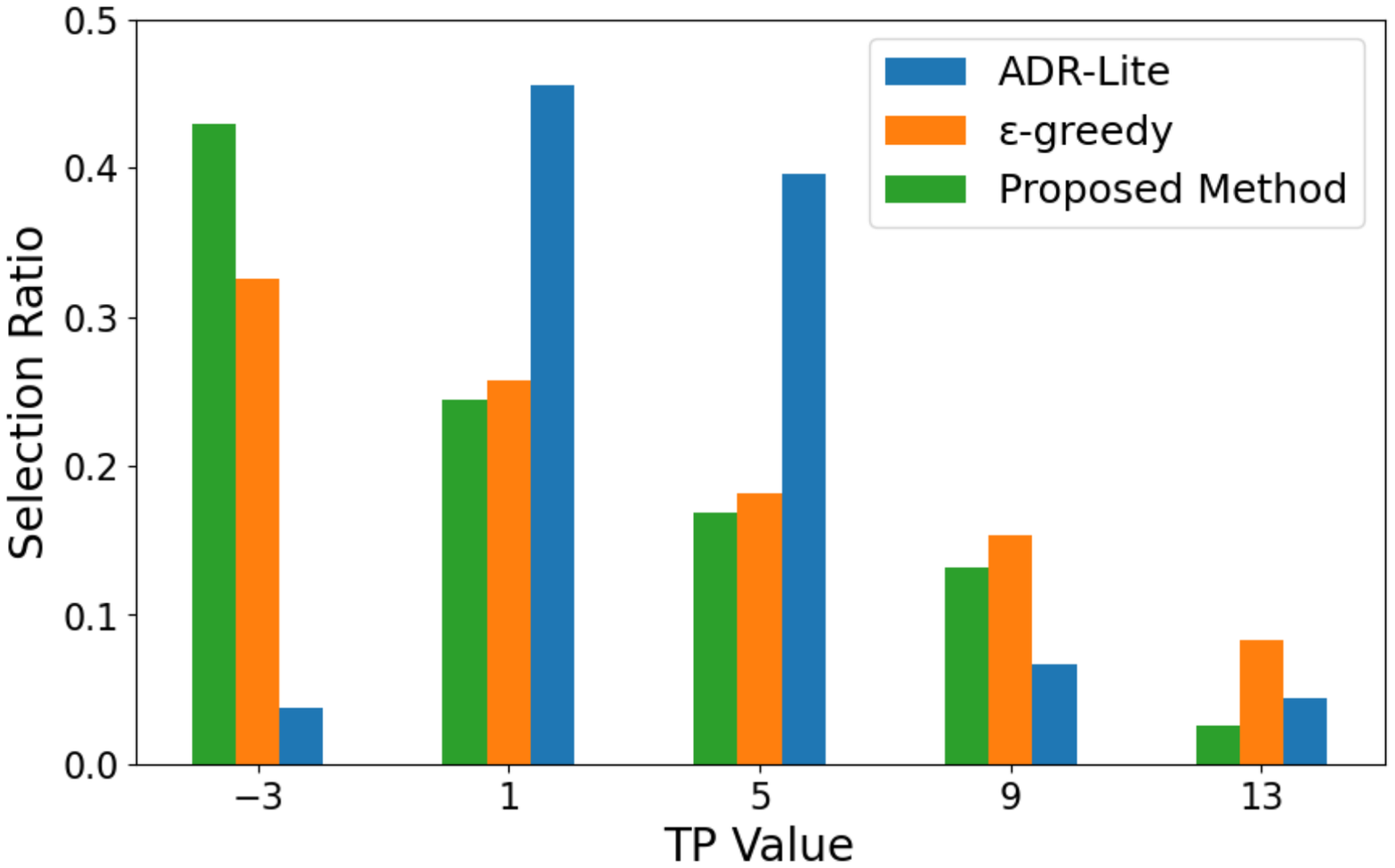}
\caption{TP Ratio}
\label{fig:TPRatio}
\end{figure}
As shown in Fig. \ref{fig:TPRatio}, our proposed method can achieve the highest proportion of selecting the minimum power. In the case of $\epsilon$-greedy, it took more time to explore compared to the proposed method, resulting in an increased probability of selecting a larger transmission power. Moreover, the ADR-Lite had the lowest proportion of selecting the minimum power. This is due to 
the tendency to select larger transmission power when a transmission fails once. 

\subsubsection{Transmission Success Rate}
Fig. \ref{fig:success} shows the transmission success rate with varying numbers of transmitters for each algorithm.\begin{figure}
\centering
\includegraphics[width=50mm, angle=270]{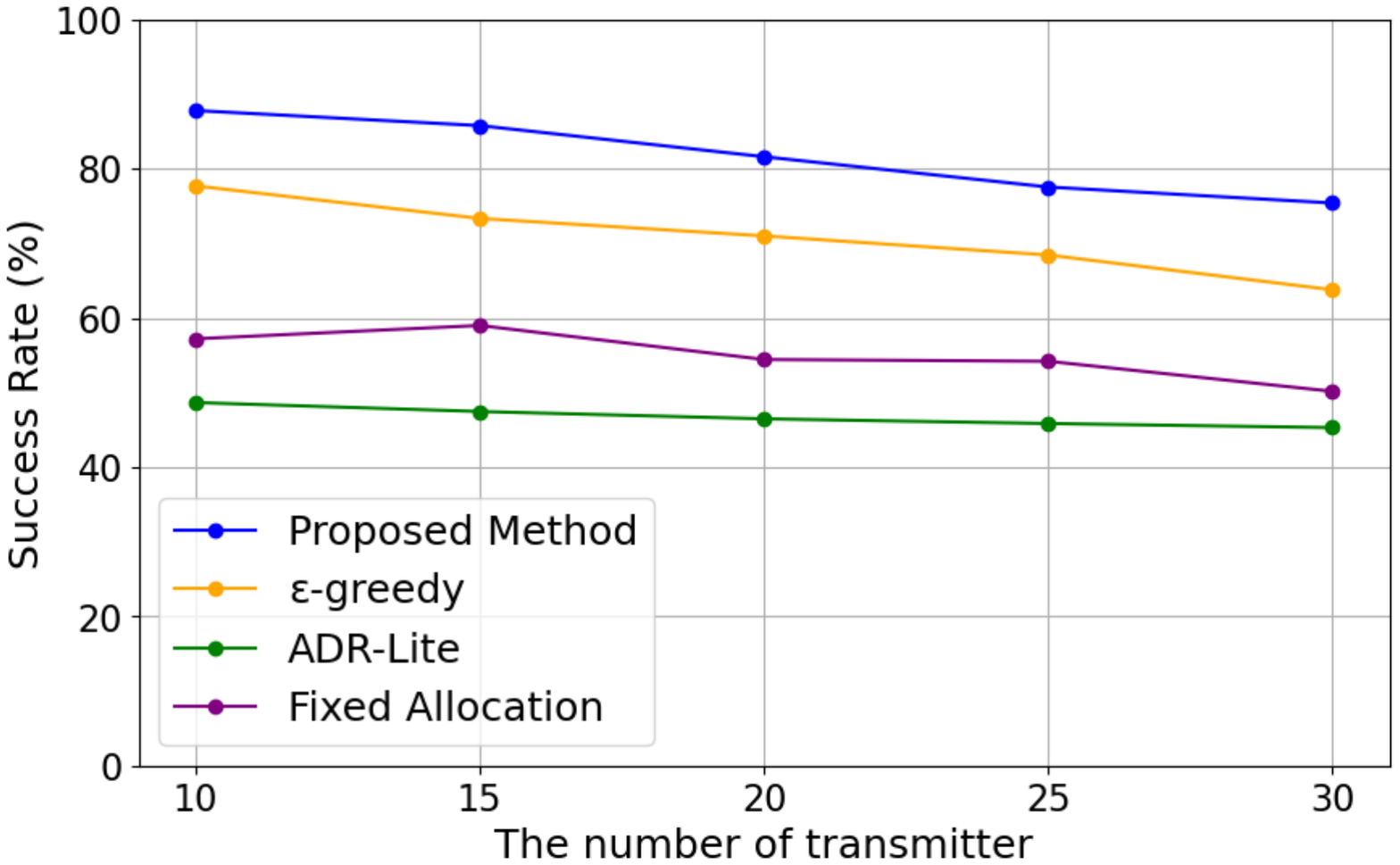}
\caption{Transmission Success Rate}
\label{fig:success}
\end{figure}
As shown in Fig. \ref{fig:success}, the transmission success rate decreases as the number of transmitters increases. This is likely due to increased traffic leading to channel congestion and communication collisions. In addition, our proposed method can achieve the highest success rate under any number of transmitters compared to other methods. Although fixed allocation evenly assigned channels, its performance was worse than that of the MAB (our proposed and $\epsilon$-greedy-based) methods. This is because it does not consider the states of other devices in the surroundings. Moreover, the ADR-Lite method cannot avoid channels with low transmission success rates as it only considers the power consumption and the results (successfully transmitted or not) of the previous transmission. 

\subsubsection{Energy Efficiency}
Fig. \ref{fig:EE} shows the results in energy efficiency with the varying numbers of LoRa EDs for each algorithm.
\begin{figure}
\centering
\includegraphics[width=50mm, angle=270]{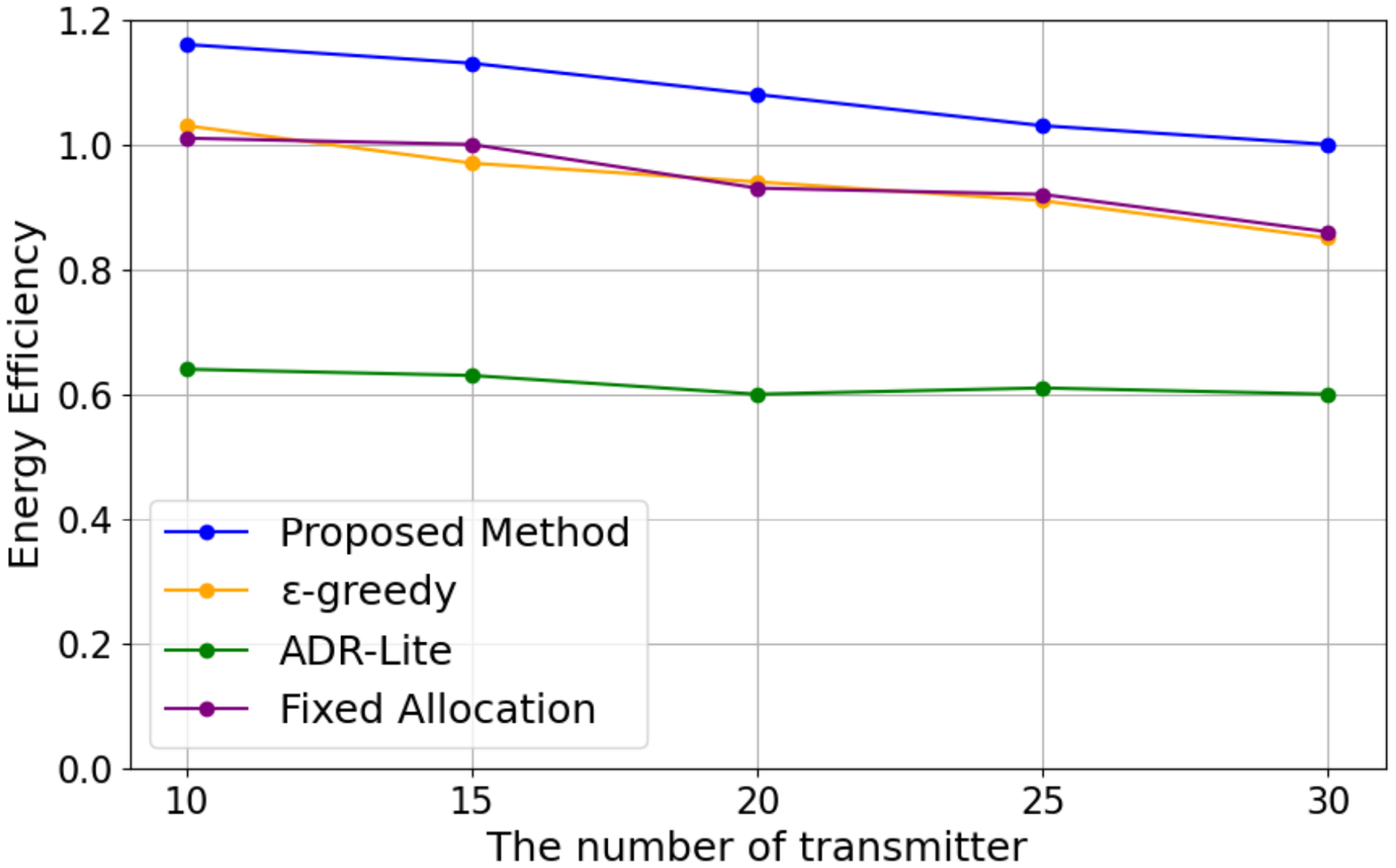}
\caption{Energy Efficiency}
\label{fig:EE}
\end{figure}
As shown in Fig. \ref{fig:EE}, the value of energy efficiency decreases as the number of transmitters increases. This is likely due to increased traffic leading to lower transmission success rates and more devices selecting larger transmission power. In addition, our proposed method can achieve the best performance under any number of transmitters, followed by the $\epsilon$-greedy method. This is because our proposed method is better at selecting lower transmission power. The ADR-Lite method performed poorly in energy efficiency because it selected larger transmission power to avoid transmission failures. 
Despite assigning the minimum transmission power, fixed allocation performed worse in energy efficiency than our proposed method. 
This is because it had a lower transmission success rate compared to our proposed method. Therefore, the trade-off between the level of the selected transmission power and the transmission success rate is also important in energy efficiency.

\section{Conclusion and Future Work}
\label{sect:conclusion}
This paper proposed an autonomous distributed transmission parameter selection method using reinforcement learning to improve the energy efficiency in LoRa networks. 
We implemented the proposed method in a constructed real high-density LoRa network and evaluated the energy efficiency and successful transmission rate performance. 
As a result, our proposed method showed the best performance in both metrics. Furthermore, the reinforcement learning-based methods performed better in both metrics compared to fixed allocation and ADR-Lite. This is considered because each LoRa ED could select smaller transmission power while avoiding channel congestion through learning. 


Future challenges include expanding the type of transmission parameters to be optimized, such as incorporating the selection of Spreading Factor (SF) and Bandwidth (BW), to further enhance energy efficiency. Additionally, increasing the number of selectable parameters or considering parameter combinations is expected to expand the search space, potentially slowing convergence. Therefore, improving the algorithm used in the initial exploration phase to efficiently handle a large search space is necessary. Furthermore, when implementing the proposed method on resource-constrained IoT devices, it is essential to conduct a more detailed analysis of the computational overhead and memory requirements to ensure feasibility.


\section*{Acknowledgment}
This work was supported by JSPS KAKENHI Grant Number 22K14263.

\end{document}